\documentclass[a4paper,10pt]{article}
\usepackage{amsmath,graphicx}

\newcommand{\pni}{\par\noindent}
\begin{document}
\title{Schwarzschild metrics, quasi-universes and wormholes}
\author{ A. G. Agnese \, and M. La Camera\\
{\em Dipartimento di Fisica dell'Universit\`a di 
Genova}\\{\em Istituto Nazionale di Fisica Nucleare, Sezione di 
Genova}\\{\em Via Dodecaneso 33, 16146 Genova, Italy}\\
{\em E-mail: agnese@ge.infn.it ; lacamera@ge.infn.it}\\}
\date{}
\maketitle 
\vspace{0.2in} 
\baselineskip = 1.1\baselineskip
\section{Introduction}
It is well known [1] that the three-dimensional space
\begin{equation}
d^{(3)}s^2 = 
\dfrac{dr^2}{1-\dfrac{r^2}{R^2}}+r^2(d\vartheta^2+\sin^2\vartheta
d\varphi^2)
\end{equation}  
of some models of closed homogeneous and isotropic universes has 
an especially simple geometry which can be seen best introducing 
a  angular coordinate \linebreak $0 \leq \chi \leq \pi$ via $r=R 
\sin\chi$ and transforming the line element (1) into the form 
\begin{equation}
d^{(3)}s^2 = R^2\,(d\chi^2 + \sin^2\chi\, d\Omega^2)
\end{equation}
where
\begin{equation}
d\Omega^2 = d\vartheta^2+\sin^2\vartheta
d\varphi^2 
\end{equation}
The metric (2) is that of a three-dimensional hypersurface of 
radius $R$ which can be represented in a flat, four-dimensional 
Euclidean embedding space.\pni
Our purpose is to employ a similar angular variable to describe 
the geometry of the exterior Schwarzschild solution and to 
investigate such a description of the interior solution also when
$\chi > \pi/2$, a possibility which appears to have been ignored 
in the literature. In this way we can introduce the concept of 
``quasi-universe'' and show that the the Einstein-Rose bridge is 
nothing else that an ``extreme wormhole connecting two 
quasi-universes''. Moreover it can be seen, in the framework of 
Brans-Dicke theory, that the Einstein-Rosen bridge becomes a 
traversable wormhole. 
\section{The exterior Schwarzschild 
solution} The exterior spherically symmetric vacuum solution, 
which by Birkhoff's theorem is also static, will be written in 
standard coordinates as \begin{equation}
ds^2 =\dfrac{dr^2}{1-\dfrac{2m}{r}} + r^2 d\Omega^2 -\, N^2(t) 
\left(1-\dfrac{2m}{r}\right) dt^2
\end{equation}
The term $N^2(t)$ allows the matching between exterior and 
interior values of $g_{tt}$ when the interior solution is not 
static and the observer is below the radius $r_1$ of the body; 
of course in the static cases $N^2(t)$ reduces to a constant. 
Such a constant shall be written as $(1-2m/r_0)^{-1}$ if the 
observer is placed at $r_0$ above the radius $r_1$; so the light 
will appear to him red-shifted if received from $r<r_0$ and 
blue-shifted if received from $r>r_0$.\pni 
Coming back to the line element (4), we want to replace the 
radial coordinate $r$ with an angular coordinate $\psi$; because 
of the covariance of Einstein's equations there are infinite ways
to accomplish the replacement. We choose to define an angular 
coordinate $\psi$ given by 
\begin{equation}
r = \dfrac{2m}{\cos^2\psi} 
\hspace{4cm} -\, \dfrac{\pi}{2} \leq \psi \leq \dfrac{\pi}{2} 
\end{equation}
when $r > 2m$, and analytically continued to
\begin{equation}
r = \dfrac{2m}{\cosh^2\psi}
\hspace{3.7cm} -\, \infty < \psi <  \infty 
\end{equation}
when $r < 2m$.
The line element (4) becomes, in the region $r > 2m$ 
\begin{equation}
ds^2 = \dfrac{16 m^2}{\cos^6 \psi}\, d\psi^2 + \dfrac{4 
m^2}{\cos^4\psi}\, d\Omega^2 -\dfrac{\sin^2\psi}{\sin^2\psi_0}\, 
dt^2
\end{equation}
Here the event horizon is placed at $\psi = 0$, while infinity is
reached at $\psi = \pm \, \pi/2$.  The metrical relations in the 
surface $t$ = constant, $\vartheta = \pi/2$ are illustrated by 
means of the surface of revolution $f(r) = \sqrt{8m(r-2m)}$ 
(remember the representation of the Flamm's paraboloid with the 
Einstein-Rosen bridge).
In the extended region $r < 2m$ one has instead the line 
element 
\begin{equation}
ds^2 = -\, \dfrac{16 m^2}{\cosh^6 \psi}\, d\psi^2 + \dfrac{4 
m^2}{\cosh^4\psi}\, d\Omega^2 
+ \dfrac{\sinh^2\psi}{\sinh^2\psi_0}\, dt^2
\end{equation}
which describes the interior of a black hole joined to the 
exterior by the event horizon placed at $\psi = 0$.
It is worth noticing that the introduction of the $\psi$ 
coordinate provides a division of the maximally 
extended Schwarzschild spacetime in four regions with two 
singularities corresponding to an equal gravitational mass, just 
as described by Kruskal-Szekeres coordinates. These 
singularities are placed at $\psi = \pm \infty$, being now 
$\psi$ a time coordinate. 
\section{The interior Schwarzschild solution} 
The gravitational field inside a celestial body, 
say a star, modelled on an ideal fluid medium with 
energy-momentum tensor \begin{equation} T_{\mu\nu} = (\rho + p) 
u_\mu u_\nu + p g_{\mu \nu} \end{equation} is given, for static 
distribution of matter and pressure and moreover under the 
hypotheses of spherical symmetry and constant mass density, by 
\begin{equation}
ds^2 = \dfrac{dr^2}{1-\dfrac{r^2}{R^2}} + r^2 d\Omega^2 - 
\left[ A - B\,\sqrt{1-\dfrac{r^2}{R^2}}\right]^2\, dt^2
\end{equation}
Here  $A$ and $B$ 
are integration constants to be determined by the matching 
conditions. We use this simple and rather unrealistic solution as
a toy model uniquely to illustrate the employ, which is new if 
$\chi > \pi/2$, of the angular coordinate $\chi$. If we now 
define this angular coordinate through 
\begin{equation}
\dfrac{r}{R} \equiv \sin\chi \hspace{6cm} 0 \leq \chi \leq \pi 
\end{equation}
the line element (10) becomes
\begin{equation}
ds^2 = R^2\, (d\chi^2 + \sin^2\chi d\Omega^2) - [A - 
B \cos\chi ]^2\, dt^2
\end{equation}
From Einstein's equations the pressure $p$ and the mass density 
$\rho$ are 
\begin{equation}
p = \dfrac{1}{8\pi R^2}\, \left[ \dfrac{3B 
\cos\chi-A}{A-B \cos\chi}\right]\ , \hspace{2.6cm} \rho = 
\dfrac{3}{8 \pi R^2} 
\end{equation}
In formulating the matching conditions to connect the exterior 
and interior Schwarzschild solutions, continuity of the metric 
and its derivatives are to be taken into account. However in
our simple example we rest on physical plausibility 
considerations, so we require that the metric is continuous 
for $\sin \chi_1=~\dfrac{r_1}{R}$, where $r_1$ is the radius 
of the body, that the pressure $p$ vanishes on its surface and 
that the observer is in the interior at an angle $\chi_0$. \pni 
As a result one obtains 
\begin{equation} \hspace{-1cm}
\sin \chi_1 = \left(\dfrac{2 m}{R}\right)^{1/3}\ 
,\hspace{3mm} A = \dfrac{3\cos\chi_1}{3\cos\chi_1 - 
\cos\chi_0} \ ,\hspace{3mm}  B = \dfrac{1}{3\cos\chi_1 - 
\cos\chi_0}
\end{equation}
where $m$ is the gravitational mass.
The line element (10) can now be written
\begin{equation}
ds^2 =  R^2 \, (d\chi^2 + \sin^2\chi 
d\Omega^2) - \left[\dfrac{3 \cos\chi_1 - \cos\chi}{3 \cos\chi_1 
- \cos\chi_0} \right]^2\, dt^2
\end{equation}
So the observer receives the frequency of light red-shifted when 
coming from inside and blue-shifted when coming from outside.
The matching to the exterior solution requires that
\begin{equation}
N^2 = \left[\dfrac{2\cos\chi_1}{3\cos\chi_1-\cos\chi_0}
\right]^2\, \left(1 - \dfrac{2m}{R\sin\chi_1}\right)^{-1} 
\end{equation}
If the observer is at the exterior the previous values of $A$ and
$B$ change accordingly.
The pressure becomes 
\begin{equation}
p = \dfrac{3}{8\pi R^2}\, \left[ \dfrac{\cos\chi - \cos\chi_1}{3 
\cos\chi_1 - \cos\chi}\right]
\end{equation}
and is obviously observer independent.
Because of definition (11) two cases are now to be considered,
depending whether for a given value of $r_1$ one chooses $\chi_1 
< \pi/2$ or $\chi_1 > \pi/2$. In the former case, while the mass 
density $\rho$ is constant, the pressure $p$, which is zero at 
the surface, increases inwards; the solution is non singular as 
long as $p$ is finite. At $r = 0$ where $p$ takes its maximum 
value, this is only possible for $\chi_1^{(1)} < \arccos\, (1/3) 
\approx 0.39 \pi$, that is, as known [2], for $r_1/(2m) > 9/8$. 
In the latter case, the pressure $p$ takes negative values in the
interior, and the solution is non singular at $r=0$ for 
$\chi_1^{(2)} > \pi/2$. In both 
cases, the weak energy condition
\begin{equation}
\rho \geq 0 \ , \hspace{1cm} \rho + p \geq 0
\end{equation}
is always satisfied. We would also point out that while the 
surface area $ S = 4 \pi R^2 \sin^2\chi_1$ is the same in the
two cases, independently of the choice made for $\chi_1$, things 
are different in calculating volumes, given by the formula
\begin{equation}
V = 4 \pi R^3\, \int_{0}^{\chi_{1}} \sin^2 
\chi \, d\chi = \pi R^3 (2 \chi_1 - \sin 
2\chi_1)
\end{equation}
To make an example let us consider two bodies having the 
same gravitational mass and the same density $\rho$ but different
values of $\chi_1$ given respectively by $\chi_1^{(1)}$ and 
$\chi_1^{(2)} = \pi - \chi_1^{(1)}$ (and so the same value of 
$\sin\chi_1$). The ratio $V^{(2)}/V^{(1)}$ of their volumes is
\begin{equation}
\dfrac{ V^{(2)}}{V^{(1)}} = \dfrac{2(\pi - \chi_1^{(1)}) + \sin 
2\chi_1^{(1)}}{2 \chi_1^{(1)} - \sin2\chi_1^{(1)}} 
\end{equation}
Therefore while the volume $V^{(1)}$ encloses a star whose matter 
is endowed by the usual properties ($\rho > 0,\ p>0$), the volume 
$V^{(2)}$ may be so large to be considered as a 
``quasi-universe'', so named because it is an universe deprived 
of a spherical void, containing  matter with 
properties ($\rho > 0,\ p < 0$, but $\rho + p > 0$); we do not 
call such a matter exotic, because it satisfies the weak energy 
condition and so also the null energy condition [3]. The 
connection between a body and a quasi-universe through a suitable
part of the Flamm paraboloid is schematically represented in 
Figure 1. A different possibility is shown in Figure 2 where now 
two quasi-universes are joined through an Einstein-Rosen bridge 
(with throat at $\psi = 0$) which can be renamed ``extreme 
wormhole''; here the matching  conditions to be fulfilled for the
second junction are the same already seen for the first, 
analogous quantities being now renamed with the same letter 
primed. Because the throat is in the vacuum, the null energy 
condition is not violated; so, according to the Morris-Thorne 
analysis [5] it is not seen as traversable by an observer placed
in a fixed forwarding station. The Einstein-Rosen bridge (or 
extreme wormhole) can also be considered as a limiting case, when
the post-Newtonian parameter $\gamma \to 1^+$, of the 
corresponding Brans-Dicke solution [6] (see in the following). 
Finally, because of the necessary equality of the gravitational 
masses in the three joined solutions, one obtains the following 
relation between the densities of the two quasi-universes $\rho$ 
and~$\rho_{1}^{'}$: 
\begin{equation} \dfrac{\rho}{\rho_{1}^{'}} =
\left(\dfrac{\sin\chi_1}{\sin\chi_{1}^{'}}\right)^2 
\end{equation}
\begin{figure}[h]
    \centering
    \begin{minipage}[t]{2.1in} 
    \fbox{ \includegraphics[width=1.9in]{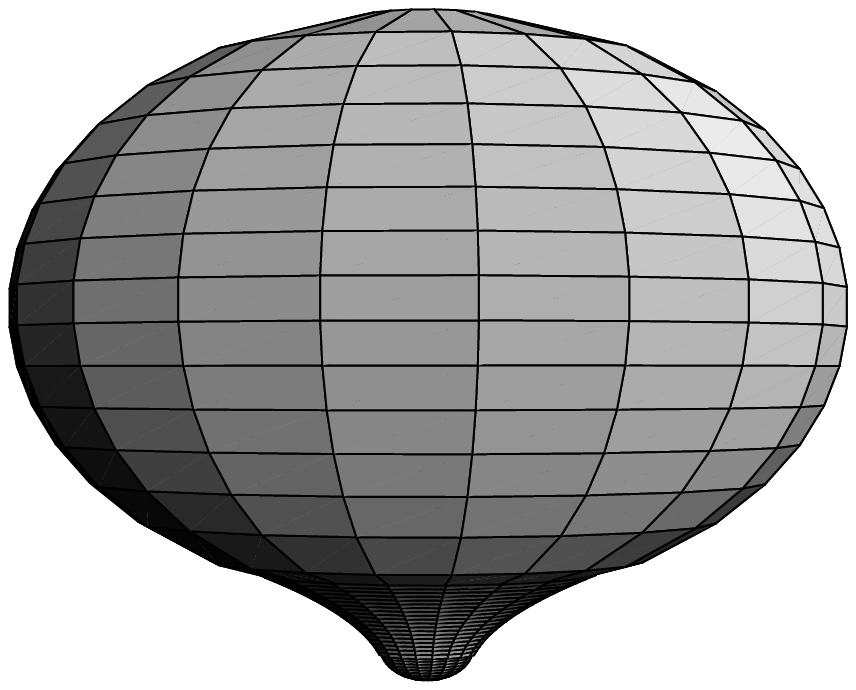}}
    \caption{The connection between a body and a quasi-universe.}
    \end{minipage} 
    \hspace{0.3in}     
    \centering
    \begin{minipage}[t]{2.1in}  
    \fbox{ \includegraphics[width=1.9in]{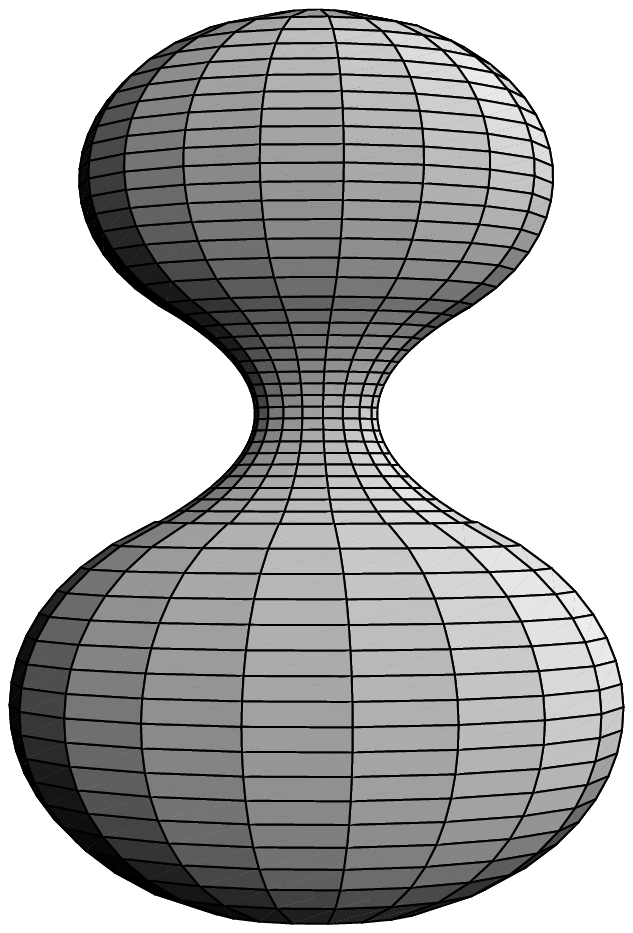}}
    \caption{The connection between two quasi-universes.}
    \end{minipage} 
\end{figure} 
\pni To summarize the above results, the metrics corresponding
to the exterior  and interior  Schwarzschild solutions have  
been rewritten replacing the usual radial coordinate with an
angular one. With respect to the exterior solution, it covers 
four different regions of the space-time. With respect to the 
interior solution, it has been extended from the case $\chi < 
\pi/2$ (first-type solution) to the case $\chi > \pi/2$ of a 
quasi-universe (second-type solution). A second-type solution can
be joined either to a first-type or to a second-type solution 
respectively through a suitable part of the Flamm's paraboloid or
through a particular Einstein-Rosen bridge (extreme wormhole) 
provided the gravitational masses are equal. \pni Let us now 
consider Equations (7) and (8) in the limiting case when the 
exterior Schwarzschild solution goes over all the remaining space
(asymptotic flatness). It is our opinion that the following 
unions ($\bigcup$) of two of the four regions - named hereafter 
$\mathit{I,II,III,IV}$ according to the customary nomenclature 
[4] - of the Kruskal-Szekeres diagram give rise to distinct 
solutions: \pni 1) $\mathit{I}\bigcup\mathit{II}$ : there is 
a singularity corresponding to a gravitational mass $m$ at $\psi 
= -\, \infty$  and a quasi-universe of 
gravitational mass $m$ and density $\rho = 0$ at the 
boundary $\psi = \pi/2$. The two regions are separated at $\psi =
0$ by an event horizon.  \pni 2) $\mathit{III}\bigcup\mathit{IV}$ 
: there is a singularity corresponding to a gravitational mass 
$m$ at $\psi = + \infty$  and a quasi-universe
of gravitational mass $m$ and density $\rho = 0$ at the 
boundary $\psi = -\,\pi/2$. The two regions are separated at 
$\psi = 0$ by an event horizon.  \pni 3) 
$\mathit{I}\bigcup\mathit{III}$ : there are two 
quasi-universes with gravitational mass $m$ and density $\rho
= 0$ at the boundaries $\psi = \pi/2$ and $\psi = -\, \pi/2$, 
connected by an extreme wormhole. 
\pni 4) $\mathit{II}\bigcup\mathit{IV}$ : the universe consists 
of two equal masses placed respectively at $\psi = -\, \infty$ 
and at $\psi = + \infty$ with a cosmological horizon at $\psi = 
0$. \pni The Penrose diagram for the maximally extended 
Schwarzschild spacetime is a representation of the set of the 
four solutions. \pni More in general, one could consider 
expanding quasi-universes, which are universes with cavities 
[7],[8],[9],[10]. Inside one of such voids there is a body whose 
inertial mass is, by the equivalence principle, equal to its own 
gravitational mass and consequently, broadening the above 
considerations,  also to the gravitational mass of the 
corresponding quasi-universe.
\section{Brans-Dicke wormholes} 
We start by considering the static 
spherically symmetric vacuum solution of the Brans-Dicke theory 
of gravitation [11]. \pni
The related calculations were performed 
by us in Ref. [6], working in the Jordan frame, where the action 
is given (in units $G_0 = c = 1$) by 
\begin{equation} 
S = \dfrac{1}{16 \pi }\,\int\, d^4x \sqrt{-\, g}\left[ \Phi R - \, 
\dfrac{\omega}{\Phi}\nabla^\alpha \Phi \nabla_\alpha \Phi \right]
\end{equation}
and with a suitable choice of gauge. Here we quote
only the results relevant for the following.\pni
The line element can be written as
\begin{equation}
ds^2 = e^{\mu(r)} dr^2 +  R^2(r) d\Omega^2 - e^{\nu(r)} dt^2
\end{equation}
where $d\Omega^2 = d\vartheta^2 + \sin^2 \vartheta d\varphi^2$ 
and, in the selected gauge:
\begin{align*}
R^2(r) = r^2 \left[1-\dfrac{2\eta}{r}\right]^{1-\gamma 
\sqrt{2/(1+\gamma)}} \tag{24a} \\ {} \\
e^{\mu(r)} = \left[ 1 - \dfrac{2\eta}{r}\right]^{-\gamma 
\sqrt{2/(1+\gamma)}}  \tag{24b} \\ {} \\
e^{\nu(r)} = \left[1 - 
\dfrac{2\eta}{r}\right]^{\sqrt{2/(1+\gamma)}} \tag{24c}
\end{align*}
\setcounter{equation}{24}
Here $\gamma$ is the post-Newtonian parameter 
\begin{equation}
\gamma = \dfrac{1+\omega}{2+\omega}
\end{equation}
and
\begin{equation}
\eta = M \sqrt{\frac{1+\gamma}{2}}
\end{equation}
Finally  the scalar field is given by
\begin{equation}
\Phi(r) = \Phi_0 \left[1-\dfrac{2\eta}{r}\right]^{(\gamma 
-1)/\sqrt{2(1+\gamma)}}
\end{equation}
while the effective gravitational coupling  $G(r)$  equals
\begin{equation}
G(r) = \dfrac{1}{\Phi(r)}\,\dfrac{2}{(1+\gamma)} 
\end{equation}
the factor $2/(1+\gamma)$ being absorbed, as in Ref. [11], in the
definition of $G$.\pni
Departures from Einstein's theory of General Relativity appear 
only if $\gamma \neq 1$, a possibility consistent with 
experimental observations  which estimate it in the range \, 
$1-0.0003 < \gamma < 1+0.0003$ corresponding to the dimensionless
Dicke coupling constant $|\omega|> 3000$.\pni 
When $\gamma <1$ and $r \to 2\eta$, then $R(r)$, $e^{\nu(r)}$ 
and $G(r)$ go all to zero. Therefore we have a singularity with 
infinite red-shift and gravitational interaction decreasing while 
approaching the singularity.\pni 
When $\gamma = 1$ exactly, one has Schwarzschild solution of 
General Relativity
\begin{equation}
ds^2 = \dfrac{dr^2}{1-\dfrac{2M}{r}}+r^2 d\Omega^2- (1-
\dfrac{2M}{r}) dt^2
\end{equation}
When $\gamma > 1$, the null energy condition $\rho + p_{j}>0$
is violated [3] and a wormhole solution is obtained [6] 
with throat at 
\begin{equation}
r_0 = \eta \left[1+\gamma\,\sqrt{\dfrac{2}{1+\gamma}}\,\right]
= M \left[ \gamma + \sqrt{\dfrac{1+\gamma}{2}}\right]
\end{equation}
to which corresponds the value $R_0$ given by equation (24a). In 
this last case beyond the throat, where $R > R_{0}$, we are faced
with two possibilities according to the value of the radial 
coordinate $r$ with respect to $r_{0}$.\pni
\begin{itemize}
\item[1) ] If $r<r_{0}$, when $r \to 2\eta$ one has $R \to 
\infty$, $g_{tt} \to 0$, $g_{rr} \to \infty$ and $G \to 
\infty$. The singularity is beyond the throat and is smeared on a
spherical surface, asymptotically large but not asymptotically 
flat. \item[2) ] If $r > r_{0}$, when $r \to \infty$ one has $R 
\to \infty$, $g_{tt} \to 1$, $g_{rr} \to 1$ and $G\to G_{N}$ (the
Newton constant), so the space is asymptotically flat. In this 
case we have a two-way traversable wormhole which, more 
generally, will be a bridge connecting two quasi-universes.
\end{itemize} 
\section{Conclusions} 
The exterior and interior Schwarzschild solutions are rewritten 
replacing the usual radial variable with an angular one. This 
allows to obtain some results otherwise less apparent or even 
hidden in other coordinate systems. In particular we have
proposed the concept of ``quasi-universe'' and described the
Einstein-Rosen bridge  as the extreme wormhole 
connecting two quasi-universes. Then we have employed the 
Brans-Dicke field to convert the non traversable Einstein-Rosen 
bridge into a traversable wormhole. There are however other 
possibilities to achieve this goal: the existence of exotic 
matter suffices for the violation of the null energy condition. 
Some other possibilities are: 
\begin{itemize} 
\item[] Scalar fields acting in the low energy limit of string 
theories. 
\item[] The Casimir energy. 
\item[] Squeezed quantum states. 
\end{itemize} 
\pni As a concluding remark, we have introduced 
wormholes connecting different universes; the possibility of 
wormholes connecting different regions of the same universe 
(called stargates in the fiction) does not seem too realistic, 
due to the large amount of exotic matter needed and the 
difficulty of its stabilization in time. 

\end{document}